\documentclass[final,3p,times,twocolumn]{elsarticle}
\usepackage{amsmath,amssymb,amsfonts,amsthm,enumerate,multirow,mathtools}
\usepackage{color}
\usepackage{siunitx}
\usepackage[linesnumbered,ruled]{algorithm2e} % for algorithms
\usepackage{diagbox} % for table
\usepackage{cleveref}
\usepackage{placeins} % for floatbarrier
\usepackage[table]{xcolor}% http://ctan.org/pkg/xcolor
\usepackage{booktabs}

\DeclareMathOperator*{\argmax}{argmax}

\usepackage{lineno}
\journal{Nuclear Instruments and Methods in Physics Research Section B}

\begin{document}

    \newcommand{\coverTitle}{Robust Inference of Two-Dimensional Strain Fields from Diffraction-based Measurements}
\newcommand{\coverAuthors}{J.N. Hendriks, C.M. Wensrich, A. Wills, V Luzin, and A.W.T Gregg}
\newcommand{\coverStatus}{Accepted for publication.}

\begin{titlepage}
    \begin{center}
        {\large \em Technical report}
        
        \vspace*{2.5cm}
        %
        %% TITLE
        {\Huge \bfseries \coverTitle  \\[0.4cm]}
        
        %
        %% AUTHORS
        {\Large \coverAuthors \\[2cm]}
        
        \renewcommand\labelitemi{\color{red}\large$\bullet$}
        \begin{itemize}
            \item {\Large \textbf{Please cite this version:}} \\[0.4cm]
            \large
            \coverAuthors. \coverTitle. \textit{Nuclear instruments and methods in physics research section B}, 444:80-90, 2019.  
        \end{itemize}
        
        %{\em \coverStatus}
        \vfill
        
        \begin{abstract}
        Diffraction-based methods have become an invaluable tool for the detailed assessment of residual strain and stress within experimental mechanics. 
        These methods typically measure a component of the average strain within a gauge volume.
        It is common place to treat these measurements as point measurements and to interpolate and extrapolate their values over the region of interest.
        Such interpolations are not guaranteed to satisfy the physical properties of equilibrium and applied loading conditions.
        In this paper, we provide a numerically robust algorithm for inferring two dimensional, biaxial strain fields over a region of interest from diffraction-based measurements that satisfies equilibrium and any known loading conditions.
        By correctly treating the measurements as gauge volume averages rather than point-wise the algorithm has better performance when large gauge volumes and subsequently shorter beam-times are used.
        This algorithm is demonstrated on simulation and experimental data and compared to natural neighbour interpolation with linear extrapolation and is shown to provide a more accurate strain field.
        \end{abstract}
        
        %           \begin{keywords}
        %               high-dimensional inference, high-dimensional particle filter, exact approximation, optimal proposal, sequential Monte Carlo, importance sampling, spatio-temporal models
        %           \end{keywords}
        
        \vfill
    \end{center}
\end{titlepage}

\begin{frontmatter}

% \title{Robust interpolation of diffraction-based strain measurements}
\title{Robust Inference of Two-Dimensional Strain Fields from Diffraction-based Measurements}

\author[NewcE]{J.N. Hendriks}
\author[NewcE]{C.M. Wensrich}
\author[NewcE]{A. Wills}
\author[ANSTO]{V Luzin}
\author[NewcE]{A.W.T Gregg}

\address[NewcE]{School of Engineering, The University of Newcastle, Callaghan NSW 2308, Australia}
\address[ANSTO]{Bragg Institute, Australian Nuclear Science and Technology Organisation (ANSTO), Kirrawee NSW 2232, Australia}

\begin{abstract}
Diffraction-based methods have become an invaluable tool for the detailed assessment of residual strain and stress within experimental mechanics. 
These methods typically measure a component of the average strain within a gauge volume.
It is common place to treat these measurements as point measurements and to interpolate and extrapolate their values over the region of interest.
Such interpolations are not guaranteed to satisfy the physical properties of equilibrium and applied loading conditions.
In this paper, we provide a numerically robust algorithm for inferring two dimensional, biaxial strain fields over a region of interest from diffraction-based measurements that satisfies equilibrium and any known loading conditions.
By correctly treating the measurements as gauge volume averages rather than point-wise the algorithm has better performance when large gauge volumes and subsequently shorter beam-times are used.
This algorithm is demonstrated on simulation and experimental data and compared to natural neighbour interpolation with linear extrapolation and is shown to provide a more accurate strain field.

\end{abstract}


\end{frontmatter}

\section{Background}

Neutron and X-ray scattering techniques provide a range of diffraction-based methods for measuring elastic strain (and hence stress) within polycrystalline solids \cite{noyan87,paranjpe2005measurement,fitzpatrick03}.  These techniques are widely applied in the assessment of residual stress within engineering components (e.g. \cite{hauk97}) and in the validation of numerical models.  They have become an invaluable tool across a wide range of areas in experimental mechanics.

\begin{figure*}
\begin{center}
    \includegraphics[width=0.9\linewidth]{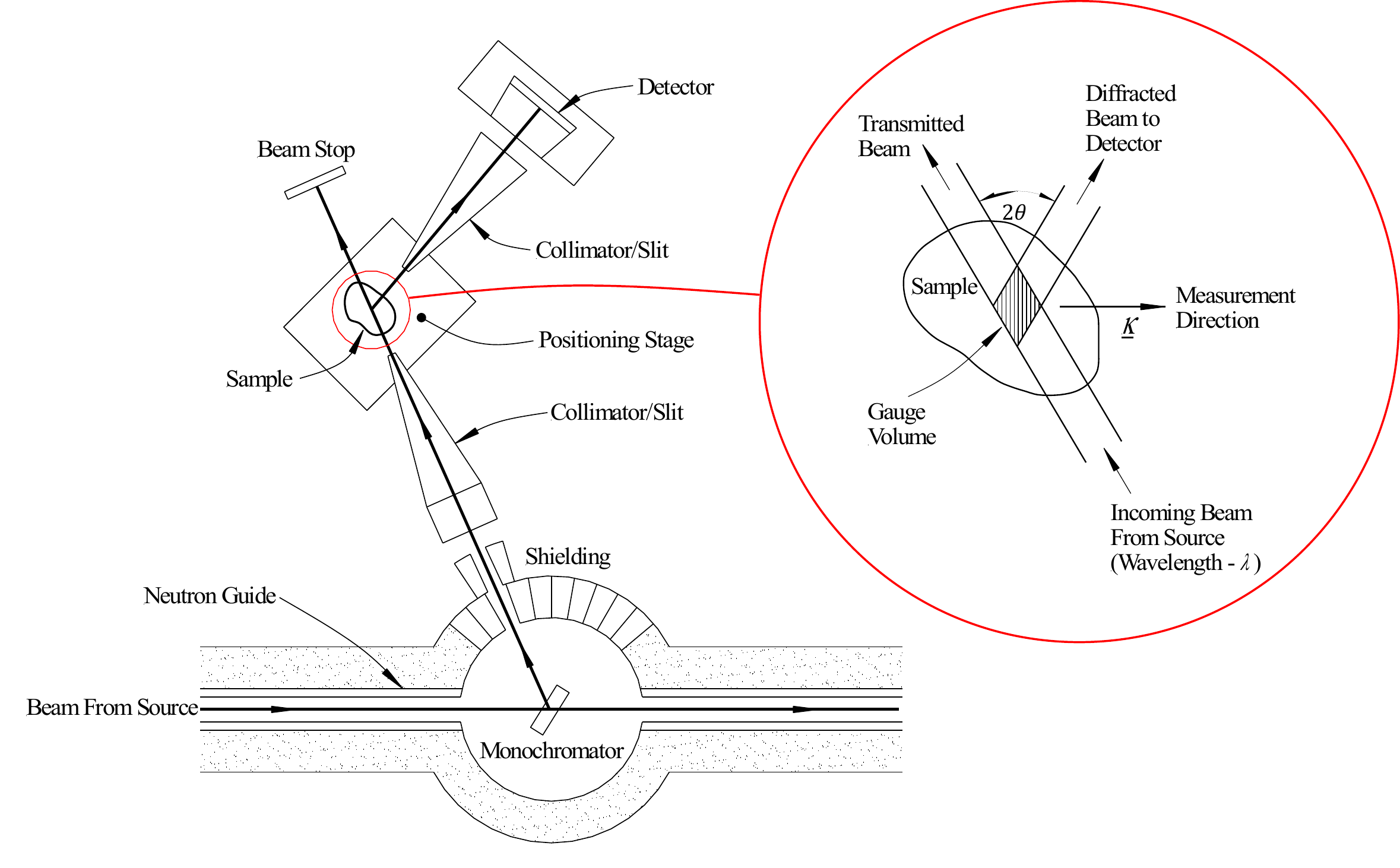}
    \caption{Typical constant wavelength strain measurement geometry (adapted from [ref]).}
    \label{fig:ConstWavelength}
\end{center}
\end{figure*}

In the vast majority of cases, these instruments measure strain in a point-wise fashion where individual components are measured as an average over a gauge volume defined by a slit or collimator system \cite{kisi2012applications}.  
For example, Figure \ref{fig:ConstWavelength} shows a typical experimental setup for constant wavelength neutron based strain measurement on instruments such as KOWARI at ANSTO \cite{kirstein2009strain,kirstein2010kowari,brule2006residual} in Australia or SALSA at ILL \cite{pirling2006salsa,pirling2006Advances,bruno2003salsa} in France.  
In this approach each strain measurement relies upon accurately measuring the Bragg angle, $\theta$, for a given wavelength, $\lambda$, from which an average lattice spacing within the gauge volume can be calculated through Bragg's law; $\lambda = 2 d \sin \theta$.  The average elastic strain within the gauge volume is then given by;
\begin{equation}\label{eq:rel_dif}
 \epsilon_\kappa = \frac{d-d_0}{d_0},
\end{equation}
where $\epsilon_\kappa = \langle \epsilon_{ij} \kappa_i\kappa_j \rangle$ is the average normal strain in the direction $\kappa$ (unit vector), $d$ is the measured lattice spacing and $d_0$ is an equivalent un-deformed reference spacing.

The full assessment of a strain field typically revolves around measuring multiple components over a mesh of points within a sample from which the full biaxial or triaxial strain distribution can be calculated and interpolated.  Given the fact that the measurement refers only to the elastic component of strain, Hooke's law can then be used to infer the distribution of the stress tensor.

The practical reality is that beam time is limited and detailed measurements can be time consuming.  This often drives a compromise in terms of the number of points that can be investigated and the size of the gauge volume used to perform these measurements.  Large gauge volumes reduce sampling times for a given level of uncertainty, however this is at the cost of spatial resolution.
Due to the averaging involved, the full amplitude of the strain field may not be captured. 

It is common practice to interpolate (and extrapolate) the strain fields from these measurements, treating the measurements as point measurements rather than as averages. 
However, there is no guarantee these interpolations will satisfy equilibrium conditions. 
Furthermore, there are often known boundary conditions (sample loading) which are not taken into account and violated by the interpolation.

Gaussian Process (GP) regression provides a non-parametric method for inferring functions from measured data.  Recently, this method was demonstrated to be an excellent approach for tomographic reconstruction of strain fields from Bragg-edge neutron transmission measurements \cite{jidling2018probabilistic}.
Through this work a GP model has been designed such that  inferred strain fields automatically satisfy equilibrium.
This approach has also been extended to include knowledge of boundary tractions in \cite{hendriks2018traction}.

In this paper, we extend the GP approach to a new problem involving the robust inference of two-dimensional strain fields from diffraction based measurements. 
This principally involves the derivation and adaptation of a different measurement model to the GP model for strain. 
We also derive a more flexible noise model for the application of traction constraints and present a robust GP algorithm that relies on QR factorisations \cite{golub96} rather than the conventional cholesky decomposition.

This new approach is demonstrated in both simulation and on experimental data where significant improvements in accuracy and/or beamtime requirements were apparent.

\section{Method}
\label{sec:method}
A detailed description of GP regression can be found elsewhere (e.g. \cite{rasmussen2006gaussian}).  To provide context in this setting, we provide a brief overview of the method in Section~\ref{sec:gaussian_processes}.
Section~\ref{sec:GP_strainfield} then summarises the GP framework for modelling strain \cite{jidling2018probabilistic} and the inclusion of traction constraints \cite{hendriks2018traction}. 
In Sections~\ref{sec:measurement_joint_distribution}, \ref{sec:approximation_method}, and \ref{sec:parameter_optimisation} we contribute a method for inferring strain fields from diffraction measurements and provide the details of a numerically robust implementation.

\subsection{GAUSSIAN PROCESS RESGRESSION} % (fold)
\label{sec:gaussian_processes}
A GP is a generalisation of the multivariate Gaussian probability distribution to a Gaussian distribution of functions, $\mathbf{f}(\mathbf{x})$. 
This distribution is uniquely defined over the spatial coordinate $\mathbf{x}$ by a mean function $\mathbf{m}(\mathbf{x}) = \mathbb{E}\left[\mathbf{f}(\mathbf{x})\right]$, and a covariance function $\mathbf{K}(\mathbf{x},\mathbf{x}') = \mathbb{E}\left[\left(\mathbf{f}(\mathbf{x})-\mathbf{m}(\mathbf{x})\right)\left(\mathbf{f}(\mathbf{x}')-\mathbf{m}(\mathbf{x}')\right)^T\right]$, where $\mathbb{E}$ refers to the expected value of the relevant quantity.

For example, we can define the distribution of a scalar function to consist of a finite series of polynomial basis functions $f(x) = \boldsymbol\phi(x)^T\mathbf{w}$, where $\boldsymbol\phi(\mathbf{x}) = (1,x,x^2,\dots,x^n)^T$ and the weights have a prior distribution $\mathbf{w}\sim\mathcal{N}(\mathbf{0},\Sigma_p)$\footnote{The notation, $\mathbf{v} ~\sim \mathcal{N}\left(\boldsymbol\mu,\Sigma \right)$ indicates that the vector $\mathbf{v}$ is a normally distributed random variable with mean $\boldsymbol\mu$ and variance $\Sigma$.}.
In this case the GP will have a prior (i.e. before the inclusion of measurements) mean and covariance function given by;
\begin{equation}\label{eq:linear_regression_model}
\begin{split}
    \mathbf{m}(x) &=  \boldsymbol\phi(x)^T\mathbb{E}[\mathbf{w}] = 0, \\
    \mathbf{K}(x,x') &= \boldsymbol\phi(x)^T\mathbb{E}[\mathbf{ww}^T]\boldsymbol\phi(x') =  \boldsymbol\phi(x)^T\Sigma_p\boldsymbol\phi(x'),
\end{split}
\end{equation}
where the prior mean function is zero as the weights had prior distributions with mean zero.  Note that the assumption of zero mean is for convenience; in any process of fitting the model to measurements, the mean and covariance functions will be updated.  If a more appropriate prior assumption is known, it can be used.

In this form, the covariance function is given by the outer product of a finite set of basis functions. 
More generally, any function that generates a positive definite symmetric covariance can be used.
A common choice is the squared-exponential;
\begin{equation}\label{eq:squared_exponential}
	K(\mathbf{x},\mathbf{x}') = \sigma_f^2\exp(-\frac{1}{2}|\mathbf{x}-\mathbf{x}'|^2),
\end{equation}
which corresponds to a Bayesian linear regression model with an infinite number of basis functions \cite{rasmussen2006gaussian}.

GP regression refers to the estimation of function values at specified inputs given a set of measurements of the form;
\begin{equation}
    y = \mathcal{L}^\mathbf{x}[\mathbf{f}(\mathbf{x})]+ b + e,
\end{equation}
where $e\sim\mathcal{N}(0,\sigma)$ represents measurement noise, $b$ is a constant bias term\footnote{The bias term $b$ can be included in the linear operator. However, here it is kept separate for clarity.}, and the measurement model $\mathcal{L}^x$ is a linear operator.
Note that, in the case where the operator contains integrals or derivatives, the superscript denotes the variable with which the operator is in respect to.

As GPs are closed under linear operators \cite{papoulis2002probability,hennig2013quasi,wahlstrom2015modeling}, a finite set of measurements $\mathbf{Y} = \{\mathbf{y}_1,\dots,\mathbf{y}_N\}$ for inputs $\mathbf{x}_1,\dots,\mathbf{x}_N$ and the function value evaluated at any given point, $\mathbf{x}_*$, are jointly Gaussian of the form;
\begin{equation*}
    \begin{bmatrix}
        \mathbf{Y} \\ \hat{\mathbf{f}}(\mathbf{x}_*)
    \end{bmatrix}\sim \mathcal{N}\left(\begin{bmatrix}\boldsymbol\mu_y \\ \mathbf{m}(\mathbf{x}_*)\end{bmatrix},
    \begin{bmatrix}
             \mathbf{K}_{\mathbf{y}\mathbf{y}'}+\Sigma_n & \mathbf{K}_{\mathbf{y}*} \\
           \mathbf{K}_{\mathbf{y}*}^T & \mathbf{K}(\mathbf{x}_*,\mathbf{x}_*)
    \end{bmatrix}
    \right)
\end{equation*} 
where $\Sigma_n=\text{diag}(\sigma_1,\dots,\sigma_n)$ is a diagonal matrix containing the variance of each measurement,
\begin{equation*}
\begin{split}
    \boldsymbol\mu_y =  \begin{bmatrix}
        \mathcal{L}^{\mathbf{x}_1}\mathbf{m}(\mathbf{x}_1) + b_1 \\ \vdots \\ \mathcal{L}^{\mathbf{x}_N}\mathbf{m}(\mathbf{x}_N) + b_N
        \end{bmatrix}, \quad 
    \mathbf{K}_{\mathbf{y}*} = \begin{bmatrix}
        \mathcal{L}^{\mathbf{x}_1}\mathbf{K}(\mathbf{x}_1,\mathbf{x}_*) \\
        \vdots \\
        \mathcal{L}^{\mathbf{x}_N}\mathbf{K}(\mathbf{x}_N,\mathbf{x}_*)
    \end{bmatrix} 
\end{split}
\end{equation*}

and
\begin{equation*}
\begingroup % keep the change local
\setlength\arraycolsep{1pt}
    \mathbf{K}_{\mathbf{y}\mathbf{y}'} = \begin{bmatrix}
        \mathcal{L}^{\mathbf{x}_1}\mathbf{K}(\mathbf{x}_1,\mathbf{x}_1)\mathcal{L}^{\mathbf{x}_1}{}^T & \cdots & \mathcal{L}^{\mathbf{x}_1}\mathbf{K}(\mathbf{x}_1,\mathbf{x}_N)\mathcal{L}^{\mathbf{x}_N}{}^T  \\
         \vdots & \ddots & \vdots \\ 
         \mathcal{L}^{\mathbf{x}_N}\mathbf{K}(\mathbf{x}_N,\mathbf{x}_1)\mathcal{L}^{\mathbf{x}_1}{}^T & \cdots & \mathcal{L}^{\mathbf{x}_N}\mathbf{K}(\mathbf{x}_N,\mathbf{x}_N)\mathcal{L}^{\mathbf{x}_N}{}^T \\ 
    \end{bmatrix}
\endgroup
\end{equation*}
A posterior estimate of $\hat{\mathbf{f}}(\mathbf{x}_*) \sim \mathcal{N}(\boldsymbol{\mu}_{\mathbf{f}_*|\mathbf{Y}},\Sigma_{\mathbf{f}_*|\mathbf{Y}})$ based on the measurements can be then written as; \cite{rasmussen2006gaussian}
\begin{equation}\label{eq:GPR}
    \begin{split}
        \boldsymbol{\mu}_{\mathbf{f}_*|\mathbf{Y}} &= \mathbf{m}(\mathbf{x}_*) + \mathbf{K}_{\mathbf{y}*}^T\left(\mathbf{K}_{\mathbf{yy}'}+\sigma_n^2 I\right)^{-1}(\mathbf{Y}-\boldsymbol\mu_y), \\
        \Sigma_{\mathbf{f}_*|\mathbf{Y}}  &= \mathbf{K}(\mathbf{x}_*,\mathbf{x}_*) - \mathbf{K}_{\mathbf{y}*}^T\left(\mathbf{K}_{\mathbf{yy}'}+\sigma_n^2 I\right)^{-1}\mathbf{K}_{\mathbf{y}*}.
    \end{split}
\end{equation}

To sumarise, the estimate of the value of a function from a set of measurements is found through the following process;

First, a prior distribution for the function is assumed and $\mathbf{m}(\mathbf{x})$ and $\mathbf{K}(\mathbf{x}_*,\mathbf{x}_*)$ calculated for the locations of interest. 
The measurement geometry is then used to calculate a prior estimate of the measurements mean $\boldsymbol\mu_y$, and the covariances $\mathbf{K}_{\mathbf{y}\mathbf{y}'}$ and $\mathbf{K}_{\mathbf{y}*}$.
Finally, these terms along with the measurement values $\mathbf{Y}$ provide a posterior estimate of the function using Equation \eqref{eq:GPR}.

Conceptually, if we were to estimate values of an unknown scalar function using GP regression with mean and covariance function given by \eqref{eq:linear_regression_model}, then this could be thought of as fitting a finite series of polynomials to the measurements.  In the case where \eqref{eq:squared_exponential} is used, the series is effectively infinite.

For a more in-depth discussion on GPs see \cite{rasmussen2006gaussian} and for details on their application to vector valued functions see \cite{wahlstrom2015modeling}.
% section gaussian_processes (end)

\subsection{GAUSSIAN PROCESS STRAIN FIELD} % (fold)
\label{sec:GP_strainfield}
A GP suitable for the estimation of two-dimensional strain fields satisfying equilibrium was shown in \cite{jidling2018probabilistic} and this method was extended to include knowledge of a sample's boundary conditions, primarily the lack of surface tractions, in \cite{hendriks2018traction}. 
A summary of this method is first provided before detailing the equations required to perform estimation from diffraction based strain measurements.

We define a GP for a distribution of Airy's Stress functions, $\varphi(\mathbf{x})$, where $\mathbf{x}=[x \ y]^T$, from which we can define a stress field as $\sigma_{xx}(\mathbf{x}) = \frac{\partial^2\varphi(\mathbf{x})}{\partial y^2}$, $\sigma_{yy}(\mathbf{x}) = \frac{\partial^2\varphi(\mathbf{x})}{\partial x^2}$, and $\sigma_{xy}(\mathbf{x}) = \frac{\partial^2\varphi(\mathbf{x})}{\partial x \partial y}$. 
Through Hooke's law, under an assumption of plane stress, this provides strain of;
\begin{equation}\label{eq:phitoeps}
    \boldsymbol{\mathcal{E}}(\mathbf{x}) =  \underbrace{\begin{bmatrix}\frac{\partial^2}{\partial y^2}-\nu\frac{\partial^2}{\partial x^2}\\ -(1+\nu)\frac{\partial^2}{\partial x \partial y} \\ \frac{\partial^2}{\partial x^2}-\nu\frac{\partial^2}{\partial y^2} \end{bmatrix}}_{\mathcal{V}^\mathbf{x}} \varphi(\mathbf{x}) = \mathcal{V}^{\mathbf{x}} \varphi(\mathbf{x}),
\end{equation}
where $\boldsymbol{\mathcal{E}} = [\epsilon_{xx}, \epsilon_{xy}, \epsilon_{yy}]^T$, and $\nu$ is poisson's ratio. 

The distribution of $\varphi(\mathbf{x})$ is assumed to have mean $m_\varphi(\mathbf{x})=0$, and covariance $K_\varphi(\mathbf{x},\mathbf{x}')$. This allows us to write a GP for the strain field with mean function $\mathbf{m}_\epsilon = \mathcal{V}^{\mathbf{x}}m_\varphi(\mathbf{x})=\mathbf{0}$ and covariance function 
\begin{equation}\label{eq:airys_map}
    \boldsymbol{K}_{\epsilon\epsilon}(\mathbf{x},\mathbf{x}') = \mathcal{V}^{\mathbf{x}} 
    K_\varphi(\mathbf{x},\mathbf{x}')
    \mathcal{V}^{\mathbf{x'}}{}^T.
\end{equation}
As stress and hence strain calculated from an Airy's stress function satisfy equilibrium, any strain field estimated based on measurements using this GP will inherently satisfy equilibrium. 
Although many options exist for the choice of covariance function, the squared-exponential is used in this work as it has previously been shown to be suitable for reconstructing strain fields in \cite{jidling2018probabilistic}. 
The shorthand $K_\varphi = K_\varphi(\mathbf{x},\mathbf{x}')$ and $\mathbf{K}_{\epsilon\epsilon} = \mathbf{K}_{\epsilon\epsilon}(\mathbf{x},\mathbf{x}')$ will be used where appropriate.

Estimation of the strain field can be improved by including any knowledge of the loading a sample is subject to. 
For example, in many cases where residual stain is of interest, there are no \textit{in-situ} loads and an assumption of zero traction can be applied to the external surfaces.
This can be achieved by incorporating artificial zero-traction measurements with no variance on the free surfaces of the sample.
As well as insuring the average strain within the sample is correct, this approach was shown to dramatically improve estimates of strain near the boundary \cite{hendriks2018traction}.

Assuming plane stress, the traction at a surface location, $\mathbf{x}_s$ can be calculated by
\begin{equation}\label{eq:traction_map}
    \mathbf{T}\hspace{-0.5mm} =\hspace{-0.5mm} \underbrace{\begin{bmatrix}
        n_1 &\hspace{-1.5mm} n_2 &\hspace{-1.5mm} 0 \\ 0 &\hspace{-1.5mm} n_1 &\hspace{-1.5mm} n_2
    \end{bmatrix}\hspace{-0.5mm}
    \frac{E}{1-\nu^2}\hspace{-0.5mm}
    \begin{bmatrix}
        1  &\hspace{-1.5mm} 0 &\hspace{-1.5mm} \nu \\
        0 &\hspace{-1.5mm} 1-\nu &\hspace{-1.5mm} 0 \\
        \nu &\hspace{-1.5mm} 0 &\hspace{-1.5mm} 1
    \end{bmatrix}}_{\mathcal{C}}\mathcal{E}(\mathbf{x}_s) \hspace{-0.5mm}= \hspace{-0.5mm}\mathcal{C}\mathcal{E}(\mathbf{x}_s)
\end{equation}
where $\mathbf{n} = [n_1 \ n_2]^T$ is the surface normal at $\mathbf{x}_s$. 
This allows the prior joint distribution of a strain $\mathcal{E}_*$ at $\mathbf{x}_*$ and a set of traction observations, $\mathbf{T}^P = [\mathbf{T}_1,\dots,\mathbf{T}_P]$ at surface points $[\mathbf{x}_{s1},\dots,\mathbf{x}_{sP}]$, to be written as

\begin{equation*}
    \begin{bmatrix}
        \mathbf{T}^P \\ \mathcal{E}_*
    \end{bmatrix} \sim 
    \mathcal{N}\left(\begin{bmatrix}
        \mathbf{0} \\ \mathbf{0}
    \end{bmatrix},\begin{bmatrix}
        \boldsymbol{K}_{TT} & \boldsymbol{K}_{T\epsilon} \\ \boldsymbol{K}_{T\epsilon}^T & \boldsymbol{K}_{\epsilon\epsilon}
    \end{bmatrix}\right),
\end{equation*}
where
\begin{equation*}
\begin{split}
    \mathbf{K}_{T\epsilon} &= \begin{bmatrix}
        \mathcal{C}\boldsymbol{K}_{\epsilon\epsilon}(\mathbf{x}_{s1},\mathbf{x}_*) \\
        \vdots \\
        \mathcal{C}\boldsymbol{K}_{\epsilon\epsilon}(\mathbf{x}_{sP},\mathbf{x}_*) 
    \end{bmatrix} \\
    \mathbf{K}_{TT} &= \begin{bmatrix}
        \mathcal{C}\boldsymbol{K}_{\epsilon\epsilon}(\mathbf{x}_{s1},\mathbf{x}_{s1})\mathcal{C}^T &\hspace{-3mm} \cdots  &\hspace{-3mm} \mathcal{C}\boldsymbol{K}_{\epsilon\epsilon}(\mathbf{x}_{s1},\mathbf{x}_{sP})\mathcal{C}^T \\
        \vdots &\hspace{-3mm} \ddots &\hspace{-3mm} \vdots \\
        \mathcal{C}\boldsymbol{K}_{\epsilon\epsilon}(\mathbf{x}_{sP},\mathbf{x}_{s1})\mathcal{C}^T &\hspace{-3mm} \cdots & \hspace{-3mm} \mathcal{C}\boldsymbol{K}_{\epsilon\epsilon}(\mathbf{x}_{sP},\mathbf{x}_{sP})\mathcal{C}^T  
    \end{bmatrix}
\end{split}
\end{equation*}

\subsection{STRAIN FIELD ESTIMATION} % (fold)
\label{sec:measurement_joint_distribution}
While traditionally measurements are considered to be of the form \eqref{eq:rel_dif}, this requires knowledge of the reference spacing $d_0$. 
In practice, this can often be problematic and involve the destruction of the sample. 
For this reason we use the lattice spacing measurements directly in our formulation of the GP.
This has the flexibility of either using a known $d_0$ or estimating its value {\color{black}(see Section~\ref{sec:parameter_optimisation})}.
The measurement model then becomes;
\begin{equation}\label{eq:diffraction_eq}
    d(\boldsymbol\eta) = \frac{d_0}{V}\int\limits_\mathcal{G}\boldsymbol\kappa^T\boldsymbol\epsilon(\mathbf{x})\boldsymbol\kappa \,\mathrm{d}V + d_0 + e,
\end{equation}
where $\boldsymbol\eta =  \{\boldsymbol\kappa,\mathcal{G}\}$, $\boldsymbol\kappa$ is the measurement direction unit vector, $\mathcal{G}$ is the gauge volume, and $e\sim\mathcal{N}(0,\sigma)$. 

In order to estimate the values of the strain field from a set of diffraction  measurements subject to traction constraints, we require the prior joint distribution of the measurements, $\mathbf{d} = [d_1,\dots,d_n]^T$, the traction observations $\mathbf{T}^P$, and the strain at the location we desire an estimate $\mathcal{E}_* = \mathcal{E}(\mathbf{x}_*)$;
\begin{equation*}
    \begin{bmatrix}
        \mathbf{d} \\ \mathbf{T} \\ \boldsymbol{\mathcal{E}}
    \end{bmatrix} \sim 
    \mathcal{N}\left(\begin{bmatrix}
        \mathbf{d}_0\\ \mathbf{0} \\ \mathbf{0}
    \end{bmatrix},\begin{bmatrix}
        \boldsymbol{K}_{\epsilon\epsilon}+\Sigma_N & \boldsymbol{K}_{dT} & \boldsymbol{K}_{d\epsilon} \\ \boldsymbol{K}_{dT}^T & \boldsymbol{K}_{TT} & \boldsymbol{K}_{T\epsilon} \\ \boldsymbol{K}_{d\epsilon}^T & \boldsymbol{K}_{T\epsilon}^T & \boldsymbol{K}_{\epsilon\epsilon}
    \end{bmatrix}\right),
\end{equation*}
where $\mathbf{d}_0$ is an $n$ by $1$ vector of the un-deformed lattice spacing.

{\color{black}For two-dimensional strain fields, three unique measurement directions---$\boldsymbol\kappa_1$, $\boldsymbol\kappa_2$, and $\boldsymbol\kappa_3$---are required in order to accurately reconstruct the cartesian strain components; $\epsilon_{xx}$, $\epsilon_{yy}$, and $\epsilon_{xy}$.
These measurement directions can be chosen by examining the mapping from cartesian strains to the normal strains in the measurement directions given by
\begin{equation*}
    \begin{bmatrix}
        \epsilon_{\kappa_1} \\ \epsilon_{\kappa_2} \\ \epsilon_{\kappa_3} 
    \end{bmatrix} =
    \underbrace{\begin{bmatrix}
        \bar{\boldsymbol\kappa}_1 \\ \bar{\boldsymbol\kappa}_2 \\ \bar{\boldsymbol\kappa}_3
    \end{bmatrix}}_{\mathbf{M}}
    \begin{bmatrix}
        \epsilon_{xx} \\ \epsilon_{xy} \\ \epsilon_{yy}
    \end{bmatrix} 
\end{equation*}
where $\bar{\boldsymbol{\kappa}} = \begin{bmatrix} \kappa_x^2 & 2\kappa_x\kappa_y & \kappa_y^2 \end{bmatrix}$. We require $\mathbf{M}$ to be invertible and hence the measurement directions are chosen to provide a well conditioned mapping.}
Measurement directions typically used are; $\boldsymbol\kappa_1 = \begin{bmatrix}1 & 0 \end{bmatrix}^T$, $\boldsymbol\kappa_2 = \begin{bmatrix}0 & 1 \end{bmatrix}^T$, and $\boldsymbol\kappa_3 = \begin{bmatrix}\frac{1}{\sqrt{2}} & \frac{1}{\sqrt{2}} \end{bmatrix}^T$.
Which give volumetric averages of $\epsilon_{xx}$, $\epsilon_{yy}$, and $\epsilon_{xy}+\frac{1}{2}(\epsilon_{xx}+\epsilon_{yy})$ respectively---allowing for easy use within more traditional point-based interpolation approaches.

The measurement model \eqref{eq:diffraction_eq} can be written in vector form as
\begin{equation}\label{eq:meaurement_map}
\begin{split}
    d(\boldsymbol\eta) &= \frac{d_0}{V}\int\limits_\mathcal{G}\bar{\boldsymbol\kappa} \mathcal{E}(\mathbf{x})\,\mathrm{d}V + d_0 + e, \\
    d(\boldsymbol\eta) &= \mathcal{L}_{\boldsymbol\eta}^\mathbf{x} \mathcal{E}(\mathbf{x}) + d_0 + e,
\end{split}
\end{equation}
where and the linear operator represents $\mathcal{L}_{\boldsymbol\eta}^\mathbf{x} = \frac{d_0 \bar{\boldsymbol\kappa}}{V}\int\limits_\mathcal{G} (\cdot)\,\mathrm{d}V$.

The required covariance terms are now given by;
\begin{equation}\label{eq:joint_covariances}
\begin{split}
    \boldsymbol{K}_{d\epsilon} &= \begin{bmatrix}
        \mathcal{L}_{\boldsymbol\eta_1}^\mathbf{x_1} \boldsymbol{K}_{\epsilon\epsilon}(\mathbf{x}_1,\mathbf{x}_*) \\
        \vdots \\
        \mathcal{L}_{\boldsymbol\eta_N}^\mathbf{x_N} \boldsymbol{K}_{\epsilon\epsilon}(\mathbf{x}_N,\mathbf{x}_*)
    \end{bmatrix}, \\
    \boldsymbol{K}_{dd} &= \begin{bmatrix}
        \mathcal{L}_{\boldsymbol\eta_1}^\mathbf{x_1} \boldsymbol{K}_{\epsilon\epsilon}(\mathbf{x}_1,\mathbf{x}_1)\mathcal{L}_{\boldsymbol\eta_1}^\mathbf{x_1}{}^T &\hspace{-3mm} \cdots &\hspace{-3mm} \mathcal{L}_{\boldsymbol\eta_1}^\mathbf{x_1} \boldsymbol{K}_{\epsilon\epsilon}(\mathbf{x}_1,\mathbf{x}_N)\mathcal{L}_{\boldsymbol\eta_N}^\mathbf{x_N}{}^T \\
        \vdots &\hspace{-3mm} \ddots &\hspace{-3mm} \vdots \\
        \mathcal{L}_{\boldsymbol\eta_N}^\mathbf{x_N} \boldsymbol{K}_{\epsilon\epsilon}(\mathbf{x}_N,\mathbf{x}_1)\mathcal{L}_{\boldsymbol\eta_1}^\mathbf{x_1}{}^T &\hspace{-3mm}\cdots &\hspace{-3mm} \mathcal{L}_{\boldsymbol\eta_N}^\mathbf{x_N} \boldsymbol{K}_{\epsilon\epsilon}(\mathbf{x}_N,\mathbf{x}_N)\mathcal{L}_{\boldsymbol\eta_N}^\mathbf{x_N}{}^T
    \end{bmatrix}, \\
    \boldsymbol{K}_{dT} &= \begin{bmatrix}
        \mathcal{L}_{\boldsymbol\eta_1}^\mathbf{x_1} \boldsymbol{K}_{\epsilon\epsilon}(\mathbf{x}_1,\mathbf{x}_{s1})\mathcal{C}^T &\hspace{-3mm} \cdots &\hspace{-3mm} \mathcal{L}_{\boldsymbol\eta_1}^\mathbf{x_1} \boldsymbol{K}_{\epsilon\epsilon}(\mathbf{x}_1,\mathbf{x}_{sP})\mathcal{C}^T \\
        \vdots &\hspace{-3mm} \ddots &\hspace{-3mm} \vdots \\
        \mathcal{L}_{\boldsymbol\eta_N}^\mathbf{x_N} \boldsymbol{K}_{\epsilon\epsilon}(\mathbf{x}_N,\mathbf{x}_{s1})\mathcal{C}^T &\hspace{-3mm}\cdots &\hspace{-3mm}\mathcal{L}_{\boldsymbol\eta_N}^\mathbf{x_N} \boldsymbol{K}_{\epsilon\epsilon}(\mathbf{x}_N,\mathbf{x}_{sP})\mathcal{C}^T
    \end{bmatrix}.
\end{split}
\end{equation}

Estimation of strain can now be performed from diffraction strain measurements subject to traction constraints using \eqref{eq:GPR}. 
However, in its present form this estimation has a prohibitive computational cost. 
This high computation time is due to the need to numerically approximate the integrals in \eqref{eq:joint_covariances}.
Except for trivial special cases where all the gauge volumes have sides in-line with the $x$ and $y$ coordinates, an analytical solution to the volume integrals is not apparent.
An alternative is to implement an approximation scheme.

% An alternative to performing numerical integrals is to implement an approximation method that allows for an analytic solution to the integrals to be calculated.

% section measurement_joint_distribution (end)

\subsection{IMPLEMENTATION} % (fold)
\label{sec:approximation_method}
An approximation scheme is implemented to reduce the computation time. 
The scheme selected allows for an analytic solution to \eqref{eq:joint_covariances} removing the need for numerical evaluation of volume integrals. 

The squared-exponential covariance function can be approximated by a finite sum of $m$ basis functions as in \cite{solin2014hilbert}; 
\begin{equation}
     K_\varphi(\mathbf{x},\mathbf{x}') =\sum_j^m \phi_j(\mathbf{x})\Sigma_{p,jj}\phi_j(\mathbf{x}')= \Phi(\mathbf{x})\Sigma_p\Phi(\mathbf{x}')^T,
\end{equation}
where each column of $\Phi(\mathbf{x})$ is a basis function $\phi_j(\mathbf{x})$ and $\Sigma_{p,jj}$ is its spectral density;
\begin{equation}\label{eq:scalar_baseis}
\begin{split}
    \phi_j(\mathbf{x}) &= \frac{1}{\sqrt{L_xL_y}}\sin(\lambda_{xj}(x+L_x))\sin(\lambda_{yj}(y+L_y)), \\
    \Sigma_{p,jj} &= \sigma_f^2 2\pi l_x l_y \exp\left(-\frac{1}{2}\left(l_x^2\lambda_{xj}^2+l_y^2\lambda_{yj}^2\right)\right).
\end{split}
\end{equation}
Loosely speaking, $L_x, L_y, \lambda_{xj}$ and $\lambda_{yj}$ control the frequency and phase of the basis functions. 
For our application they were chosen such that the basis functions spanned a region where their spectral densities, $\Sigma_{pjj}$, were greater than a minimum threshold.
This helps to ensure that the dominant frequencies of the response are captured while maintaining numerical stability.

To apply this approximation to the problem of estimating strain from diffraction measurements we are required to map the basis functions through \eqref{eq:airys_map}, \eqref{eq:traction_map}, and \eqref{eq:meaurement_map} giving the matrices; 
\begin{equation*}
\begin{split}
    &\Phi_{*,j}=\phi_{\epsilon,j}(\mathbf{x}_*) = \mathcal{V}^{\mathbf{x}}\phi_j(\mathbf{x}_*), \\
    &\Phi_{d,ij} = \phi_{d,j}(\boldsymbol{\eta}_i) = \mathcal{L}_{\boldsymbol\eta_i}^{\mathbf{x}} \phi_{\epsilon,j}(\mathbf{x}) \qquad \qquad i = 1,\dots,N,\\
    &\Phi_{T,kj} = \phi_{T,j}(\mathbf{x}_{s,k}) = \mathcal{C} \phi_{\epsilon,j}(\mathbf{x}_{s,k}) \qquad k = 1,\dots,P,
\end{split}
\end{equation*}

The evaluation of these basis functions is found in \ref{app:approx_method}. 
Instead of forming the full joint prior distribution and calculating an estimate using \eqref{eq:GPR}, which scales badly with $N+P$ \cite{rasmussen2006gaussian}, an estimate of $\mathcal{E}_* = \mathcal{E}(\mathbf{x}_*)$ and its variance can be calculated by
\begin{equation*}
    \begin{split}
        \boldsymbol\mu_{\boldsymbol\epsilon_* | \mathbf{Y}} &= \Phi_*\left(\Phi_Y(\mathbf{x})^T\Sigma_Y^{-1}\Phi_Y(\mathbf{x})+\Sigma_p^{-1}\right)^{-1}\Phi_Y(\mathbf{x})^T\Sigma_Y^{-1}(\mathbf{Y}-\boldsymbol\mu_y) \\
         \boldsymbol\Sigma_{\boldsymbol\epsilon_* | \mathbf{Y}} &= \Phi_*\left(\Phi_Y(\mathbf{x})^T\Sigma_Y^{-1}\Phi(\mathbf{x})+\Sigma_p^{-1}\right)^{-1}\Phi_*^T,
    \end{split}
\end{equation*}
where $\Phi_y = [\Phi_d^T \ \Phi_T^T]^T$, $\mathbf{Y}=[\mathbf{d}^T \ (\mathbf{T}^P)^T]^T$ and $\Sigma_y = \text{diag}(\sigma_1,\dots \sigma_N, \mathbf{0}_{1\times P})$, $\boldsymbol\mu_y = [\mathbf{d}_{0,1\times N} \ \mathbf{0}_{1\times P}]^T$, and $\Sigma_{n} = \text{diag}(\sigma_i,\dots,\sigma_n)$. 
This reduces the complexity of the regression from $\mathcal{O}\left((N+P)^3\right)$ to $\mathcal{O}\left((N+P)m^2\right)$ \cite{rasmussen2006gaussian}.
In contrast to the equations given in \cite{rasmussen2006gaussian,solin2014hilbert} this does not assume all measurements to have the same variance which is necessary for the inclusion of traction constraints as traction 'measurements' have zero variance.

A robust implementation of these equations is found in Algorithm~\ref{alg:compute}, where the calculation of of $\Sigma_n^{-\frac{1}{2}}$ and $\Sigma_p^{-\frac{1}{2}}$ is trivial as both matrices are diagonal.
The use of QR factorisation instead of the Cholesky algorithm more routinely used provides greater numerical stability and does not require $\Phi_Y(\mathbf{x})^T\Sigma_Y^{-1}\Phi_Y(\mathbf{x})+\Sigma_p^{-1}$ to be positive definite.
This is important during parameter optimisation during which particular parameter choices could cause the equations to become ill-conditioned.
Ameliorating the ill-conditioning of this matrix affords for a larger region of attraction for the optimisation problem.

\begin{algorithm}
    \SetKwInOut{Input}{Input}
    \SetKwInOut{Output}{Output}

    % \underline{function Euclid} $(a,b)$\;
    \Input{$\Phi_*$(Estimation basis functions),$\mathbf{Y}$ (measurements), $\boldsymbol\mu_y$ (prior measurement mean), $\Phi_Y$ (Measurement basis functions), $\Sigma_y$ (measurement variance), \\$\Sigma_p$ (basis function spectral densities)} 
    $A := \begin{bmatrix}
        \Sigma_y^{-\frac{1}{2}}\Phi_y(\mathbf{x}) \\
        \Sigma_p^{-\frac{1}{2}}
    \end{bmatrix}$ \\
    $R : = \text{qr}(A)$\\
    $C_{ij} := R_{ij} \quad \forall i=1,\dots,m, \forall j=1,\dots,m$\\
    $\mathbf{\alpha} := C^{-1} (C^{-T} (\Phi_y(\mathbf{x})^T\Sigma_y^{-1}(\mathbf{Y}-\boldsymbol\mu_y)))$\\
    $\boldsymbol\mu_{\boldsymbol\epsilon_* | \mathbf{Y}} := \Phi_* \alpha$ \\
    $\boldsymbol\Sigma_{\boldsymbol\epsilon_* | \mathbf{Y}} := \Phi_* (C^{-1}  (C^{-T} \Phi_*^T))$\\
    \caption{Compute mean and variance}\label{alg:compute}
\end{algorithm}

% section approximation_method (end)

\subsection{PARAMETER OPTIMISATION} % (fold)
\label{sec:parameter_optimisation}
The squared-exponential covariance function is governed by the parameters $\boldsymbol\theta = \{l_x,l_y,\sigma_f\}$. 
These parameters can be determined by an optimisation process where we maximise the marginal log likelihood of the measurements \cite{rasmussen2006gaussian}:
\begin{equation*}
    \boldsymbol\theta_* = \argmax_\theta \left[ -\frac{1}{2}\log\det(\mathbf{K}_{dd} + \Sigma_n) - \frac{1}{2}\mathbf{d}^T(\mathbf{K}_{dd} + \Sigma_n)^{-1}\mathbf{d}\right]
\end{equation*}
When the approximation shown in Section~\ref{sec:approximation_method} is used $\mathbf{K}_{dd}$ is replaced with $\Phi_d \Sigma_p \Phi_d^T$ and the expressions for the log likelihood and its derivatives are given in \ref{sec:optim_expressions}. 
The parameters can then be determined using a suitable gradient-based method such as the BFGS algorithm in \cite{wright1999numerical}.

Additionally, $d_0$ can be included in the set of parameters that are optimised. 
This requires that we include the traction observations into the marginal log-likelihood.
{\color{black}As the average strain within the sample is defined by the sample's boundary tractions (see \cite{cristescu2003mechanics}) and variations in $d_0$ alter the average strain of the reconstuction, it follows that the maginal log likelihood will be maximised for the true $d_0$ value. 
}
% To avoid the the optimisation finding a local maxima, which would result in an incorrect $d_0$ parameter and subsequent systematic error, multiple initial $d_0$ parameters close to the materials nominal reference spacing should be trialed.
% Estimated $d_0$ values were found to converge to the true value in simulation and the measured value for experimental data.

% section parameter_optimisation (end)

% section method (end)

\section{RESULTS}
\label{sec:results}

\subsection{SIMULATION}
\begin{figure}[!ht]
    \centering
    \includegraphics[width=0.8\linewidth]{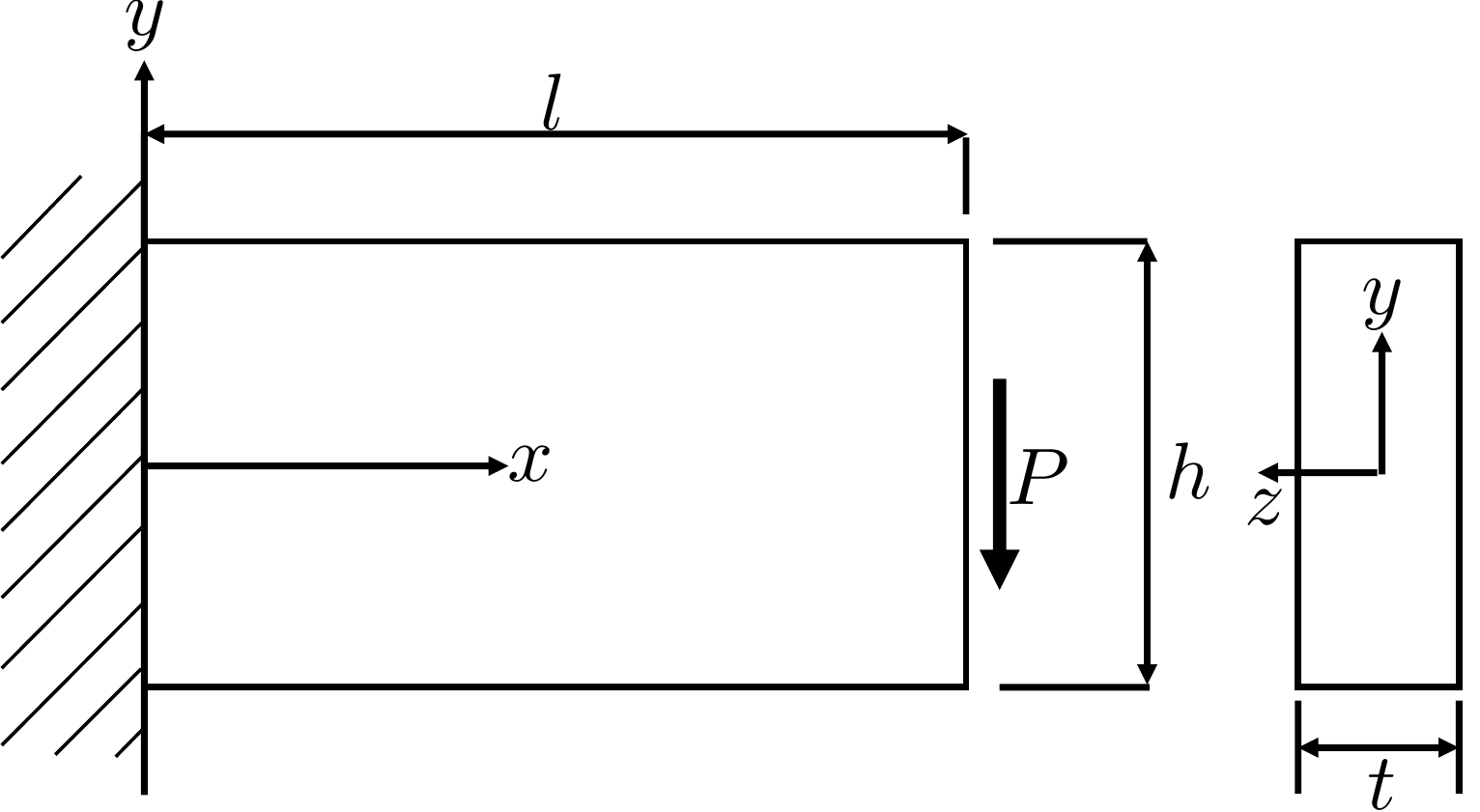}
    \caption{Cantilever beam geometry and coordinate system with $l = 20\si{mm}$, $h=10\si{mm}$, $t=5\si{mm}$, $E = 200\si{\giga\pascal}$, $P = 2\si{\kilo\newton}$, $\nu = 0.3$, and $I = \frac{th^3}{12}$.}
    \label{fig:CB_geom}
\end{figure}

The classical cantilevered beam (see Figure~\ref{fig:CB_geom}) is used to demonstrate the method and provide a comparison to standard interpolation.
Assuming plane stress, the Saint-Venant approximation to the strain field is \cite{beer2010mechanics}:
\begin{equation}
    \boldsymbol{\mathcal{E}}(\mathbf{x}) = \begin{bmatrix}
        \frac{P}{EI}(L-x)y  \\
        -\frac{(1+\nu)P}{2EI}\left(\left(\frac{h}{2}\right)^2-y^2\right) \\
        -\frac{\nu P}{EI}(L-x)y
    \end{bmatrix}
\end{equation}

Measurements were simulated from this strain field at 24 points using \eqref{eq:meaurement_map} with the standard measurement directions $\boldsymbol\kappa_1$,$\boldsymbol\kappa_2$, and $\boldsymbol\kappa_3$. 
A gauge volume size of $2\times\SI{2}{\milli\meter}$ was used and the measurements were corrupted by zero-mean Gaussian noise of standard deviation $\sigma_m = 5\times 10^{-5}$.

\begin{figure*}
\begin{center}
    \includegraphics[width=0.7\linewidth]{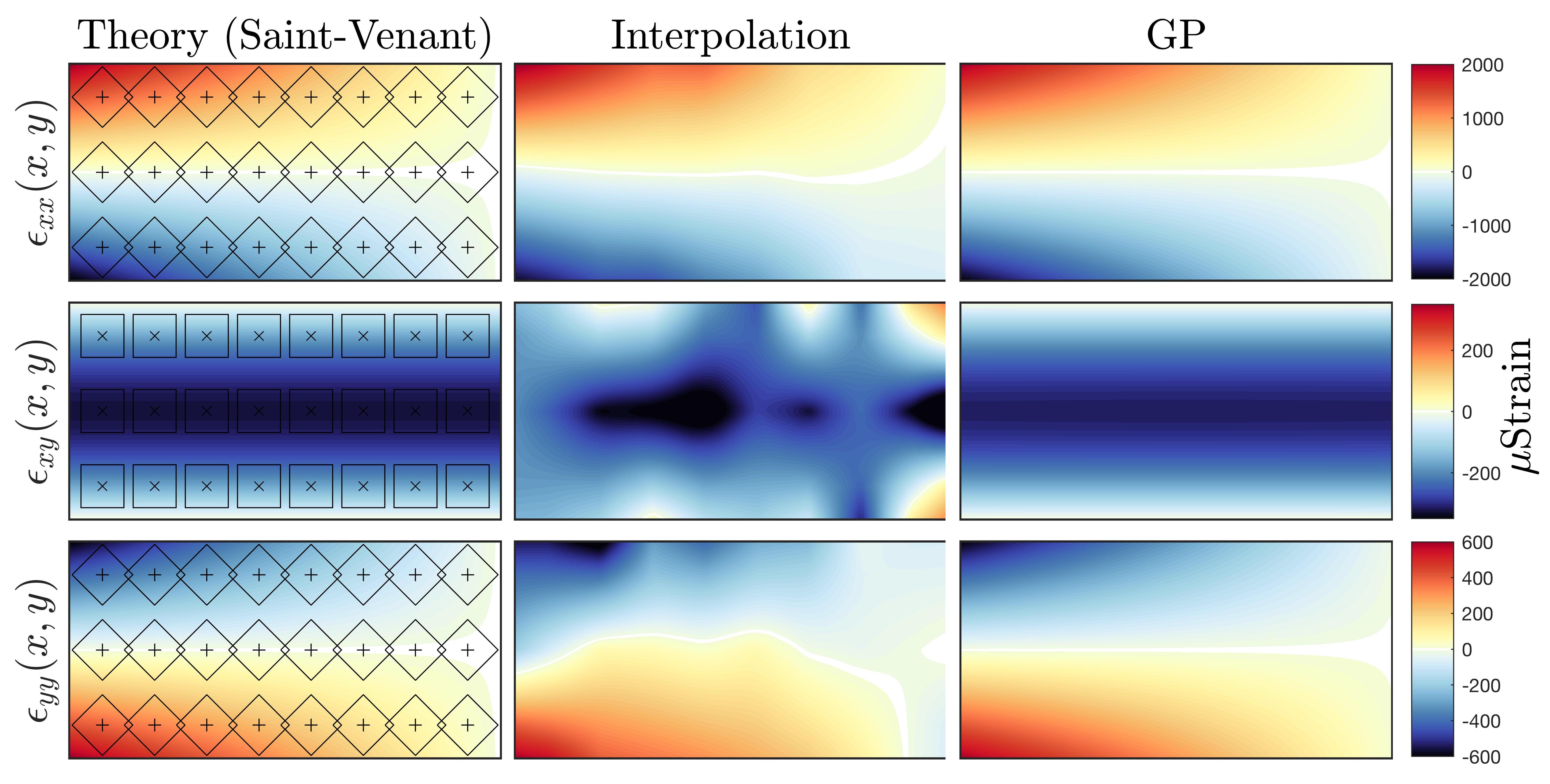}
    \caption{(Left) Theory (Saint-Venant) strain field with measurement gauge volumes. (Centre) Natural neighbour interpolation and linear extrapolation using MATLABS's \texttt{scatteredInterpolant} function. (Right) strain fields inferred using GP regression method.}
    \label{fig:CB_results}
\end{center}
\end{figure*}

The strain fields resulting from a natural neighbour interpolation (which is $C^1$ continuous) and linear extrapolation of the measurements using MATLAB's \texttt{scatteredInterpolant} function are shown in Figure~\ref{fig:CB_results} alongside results inferred through the GP method.
In the GP method, Zero-traction constraints were enforced at 100 points along the top and bottom of the beam where there is no external loading.
In the $\epsilon_{xx}$ component the interpolation has a relative error of 29.3\% while the GP has a relative error of 12.7\%. 
Over all components the interpolation has a relative error of 5.1\% while the GP has a relative error of 2.6\%.

\begin{figure}
\begin{center}
    \includegraphics[width=0.8\linewidth]{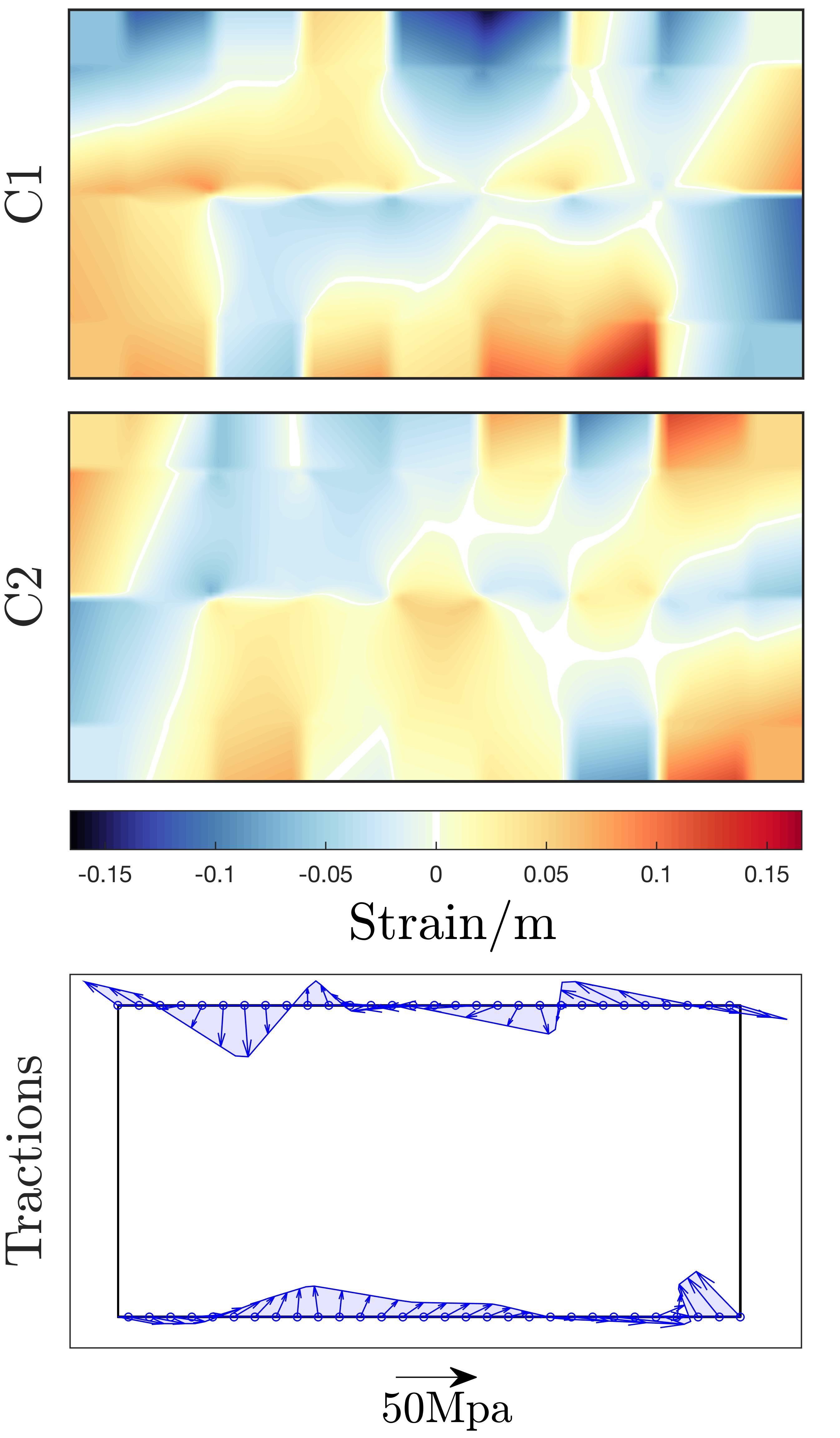}
    \caption{Violation of physical properties by the Interpolated strain field. (Top and Centre) Equilibrium constraint violations; (Bottom) Equivalent boundary tractions calculated along the top and bottom of the beam.}
    \label{fig:CB_violations}
\end{center}
\end{figure}

A measure of how well the interpolated strain field represents a physical strain field can be provided using the equilibrium constraints. 
Assuming plane stress, we can write the equilibrium constraints as;
\begin{equation}\label{eq:equi_constraints}
\begin{split}
C_1 &= \frac{\partial}{\partial x}(\epsilon_{xx}+\nu\epsilon_{yy})+\frac{\partial}{\partial y}(1-\nu)\epsilon_{xy} = 0 \\
C_2 &= \frac{\partial}{\partial x}(1-\nu)\epsilon_{xy} + \frac{\partial}{\partial y}(\epsilon_{yy}+\nu\epsilon_{xx}) = 0
\end{split}
\end{equation}
$C_1$ and $C_2$ were calculated for the interpolated strain field using a central difference approximation and are shown in Figure~\ref{fig:CB_violations}. 
Significant violation of these constraints can be observed.
In contrast, the strain field inferred using the GP method automatically satisfies these constraints.

The feasibility of the interpolated strain field can also be assessed by calculating the equivalent tractions along the top and bottom of the beam. 
The resulting tractions are shown in Figure~\ref{fig:CB_violations}.
These tractions should be zero, however, the interpolated strain field has equivalent tractions with magnitudes up to $42$Mpa.
As before, the GP strain field has been constrained to satisfy this condition, whereas it is not possible to apply these constraints to conventional interpolation methods, such as MATLAB's \texttt{scatteredInterpolant}.

\subsection{EXPERIMENTAL} % (fold)
\begin{figure}[!ht]
    \centering
    \includegraphics[width=0.32\linewidth]{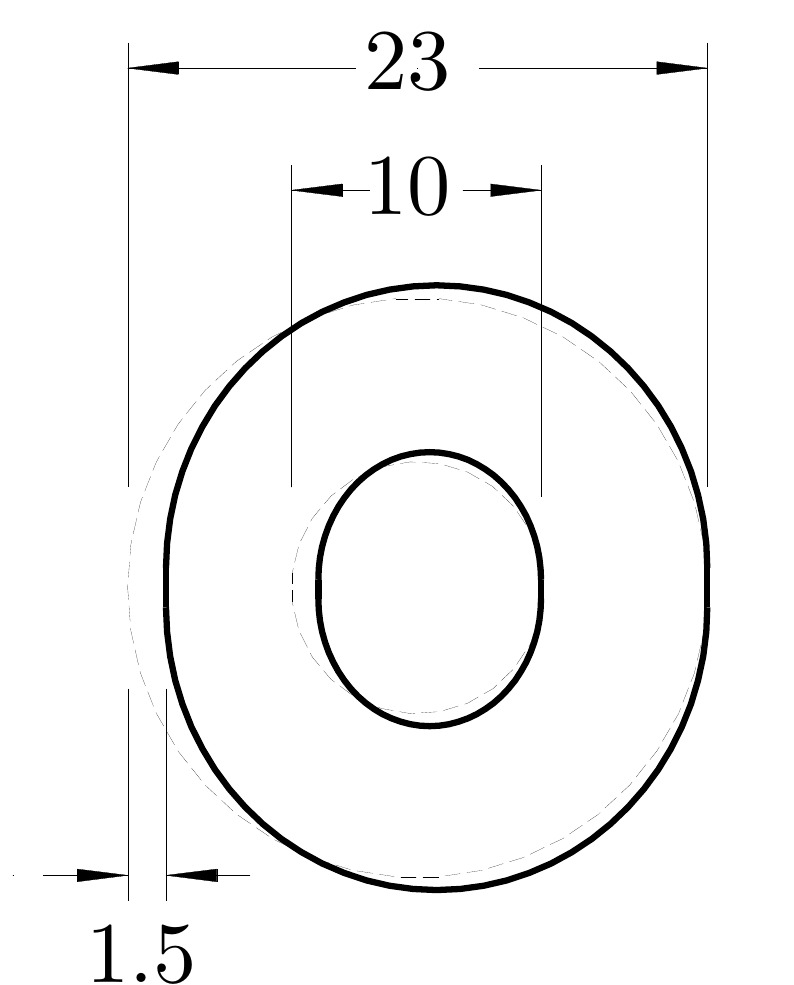}
    \includegraphics[width=0.64\linewidth]{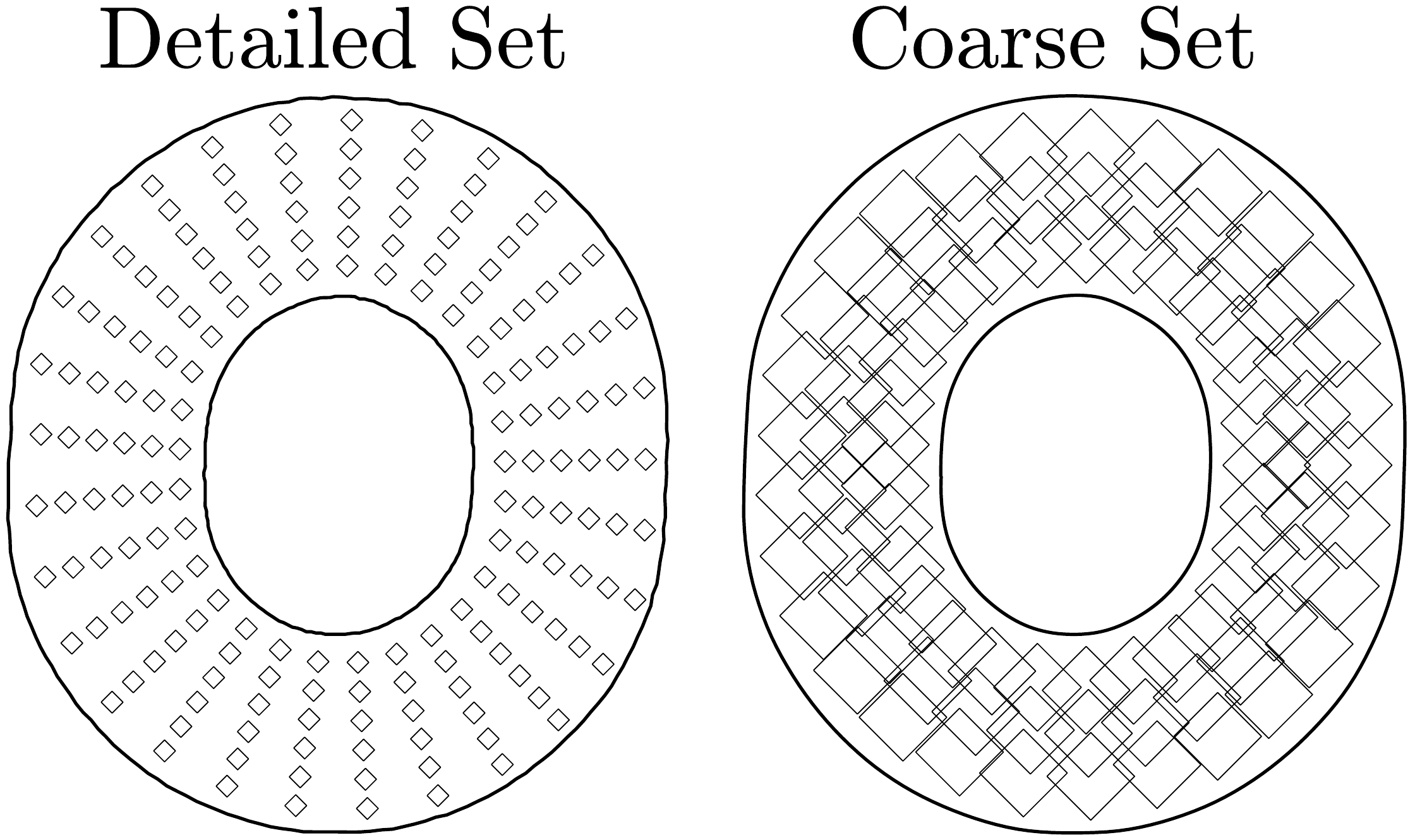}
    \caption{(Left) The crushed ring sample geometry (dimensions are in mm) and (Centre and Right) experiment gauge volumes for the two measurement sets. Gauge volumes are shown for measurement directions $\boldsymbol\kappa = [1 \, 0]^T$ and $\boldsymbol\kappa = [0 \, 1]^T$, for measurement direction $\boldsymbol\kappa=\left[\frac{1}{\sqrt{2}} \, \frac{1}{\sqrt{2}}\right]^T$ the gauge volumes are rotated by $45^\circ$.}
    \label{fig:sample_geometry}
\end{figure}

\label{sec:experimental}
The GP regression method was applied to measurements from the KOWARI constant wavelength strain-diffractometer at the Australian Centre for Neutron Scattering within the Australian Nuclear Science and Technology Organisation (ANSTO) \cite{kirstein2009strain,kirstein2010kowari,brule2006residual}.
Diffraction strain measurements were taken from the residual strain field within an EN26 steel crushed ring formed by plastically deforming a hollow cylinder.
This sample was initially heat treated to relieve residual stress, giving a final hardness of $\SI{290}{HV}$. 
It was then plastically deformed by $\SI{1.5}{mm}$ using approximately $\SI{8.4}{\kilo\newton}$ of load as shown in Figure~\ref{fig:sample_geometry}.

Two sets of measurements were made; a detailed and a coarse set, again as shown in Figure~\ref{fig:sample_geometry}.
Both sets were based on the relative movement of the $(211)$ diffraction peak measured using neutrons of wavelength $\lambda = 1.\SI{67}{\text{\AA}}$ (i.e. $90^\circ$ geometry). For the detailed set, 174 measurement locations and a gauge volume of $0.5\times 0.5\times \SI{14}{\milli\meter}$ was used. 
For the coarse set, 87 measurement locations and a gauge volume of $2\times 2\times \SI{14}{\milli\meter}$ was used.
{\color{black}For both sets, three measurement directions were used at each location; $\kappa_1 = \left[1,0,0\right]^T$, $\kappa_2 = \left[0,1,0\right]^T$, and $\kappa_3 = \left[\frac{1}{\sqrt{2}},\frac{1}{\sqrt{2}},0\right]^T$.
These directions were chosen so that the measured volumetric strains could be easily mapped to the strain components for the more traditional point-based interpolation approach; $\epsilon_{xx} = \epsilon_{\kappa_1}$, $\epsilon_{yy} = \epsilon_{\kappa_2}$, and $\epsilon_{xy} = \epsilon_{\kappa_3} - \frac{1}{2}\left(\epsilon_{\kappa_1} + \epsilon_{\kappa_2}\right)$.}
Sampling times were based on achieving measurement uncertainty with standard deviation $\sigma_m = 7\times 10^{-5}$.
This required approximately fifteen hours per measurement direction for the detailed set and approximately 1.25 hours per measurement direction for the coarse set.

\begin{figure}
\begin{center}
    \includegraphics[width=0.8\linewidth]{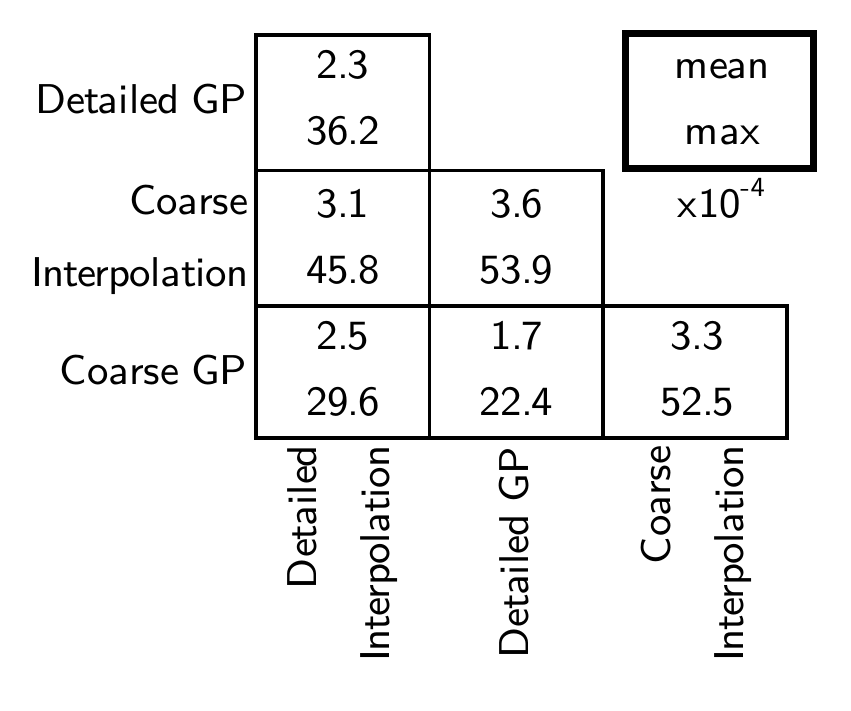}
    \caption{Comparison of the absolute difference between the strain fields, $\Delta = |\mathcal{E}_1 - \mathcal{E}_2|$. Strain fields generated from the detailed and coarse sets using both natural neighbour interpolation (Interp) and the GP method are compared. The mean and maximum of the absolute difference are provided. All values are $\times 10^{-4}$. }
    \label{tab:rel_errors}
\end{center}
\end{figure}

\begin{figure*}
\begin{center}
    \includegraphics[width=0.75\linewidth]{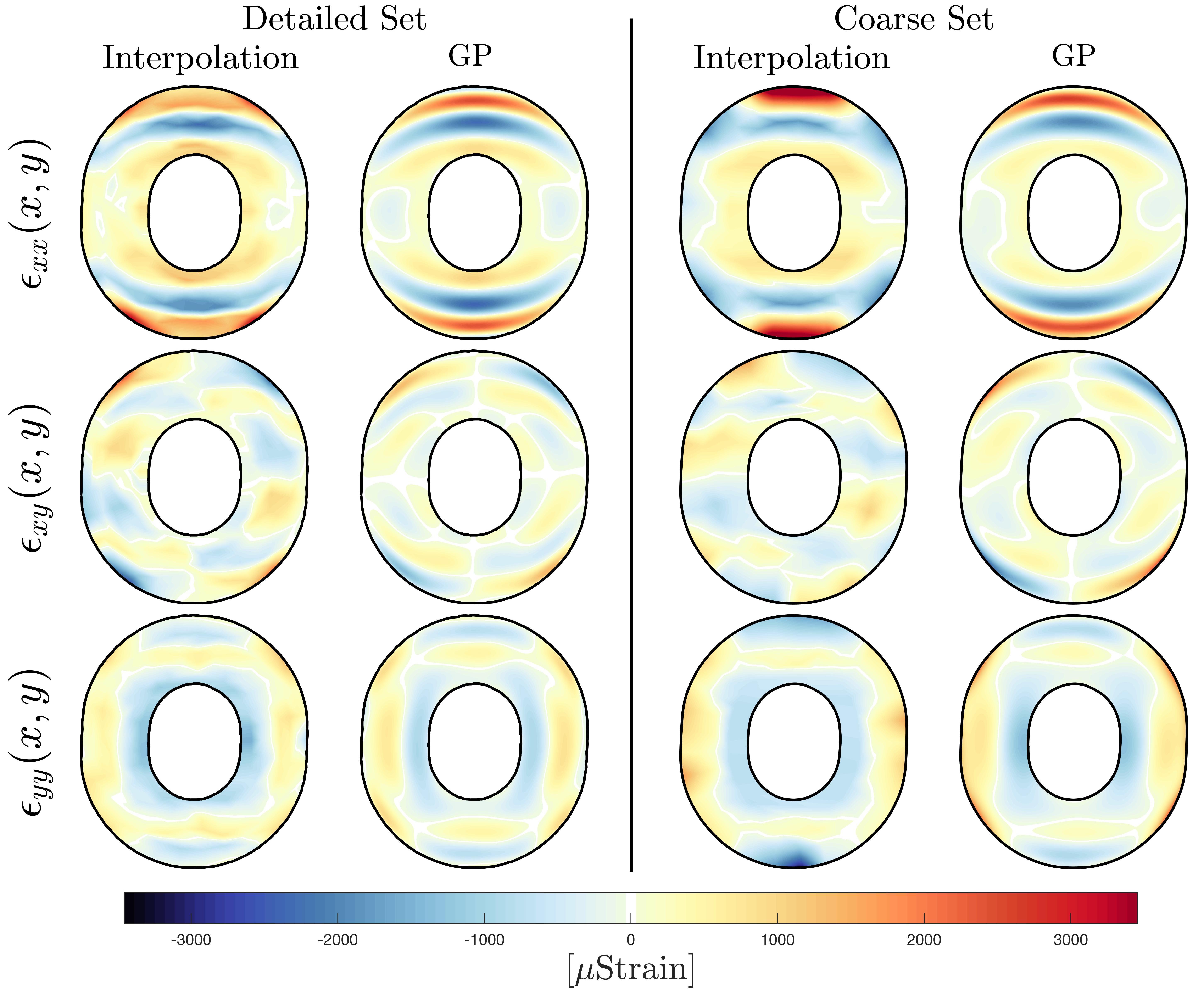}
    \caption{Strain fields generated from both experimental measurement sets using natural neighbour interpolation with linear extrapolation compared to the GP method. (Left) Detailed measurement set containing measurements at 174 locations with a gauge volume of $0.5\times0.5\times\SI{14}{\milli\meter}$. (Right) Coarse measurement set containing measurements at 87 locations with a gauge volume of $2\times2\times\SI{14}{\milli\meter}$. }
    \label{fig:CRing_results}
\end{center}
\end{figure*}

Strain fields generated using the MATLAB \texttt{scatteredInterpolant} intrinsic function and the GP method are shown in Figure~\ref{fig:CRing_results}. 
For the GP method, zero-traction measurements were included at 200 points on both the interior and exterior boundaries.
Although the two approaches only provide an interpretation of the KOWARI measurements, the strain fields generated from the detailed set should be considered closer to the truth than those generated from the coarse set.
Table~\ref{tab:rel_errors} provides a comparison of the mean of the difference between these fields.
Regardless of which method we compare to, the strain field generated from the coarse set using the GP method shows closer agreement to the detailed set than the interpolated strain field.
In particular, for the $\epsilon_{xx}$ component the \texttt{scatteredInterpolant} function has created concentrations of tension that extend to the top and bottom boundary rather than the arc of tension slightly removed from the boundary seen in the detailed sets.

If we consider the detailed strain field inferred from the GP to be the closest to the truth, then Table~\ref{tab:rel_errors} would suggest that the GP method allows us to estimate the strain field from the coarse set at least as well as a nearest neighbour interpolation from the detailed set.

\begin{figure*}
\begin{center}
    \includegraphics[width=0.55\linewidth]{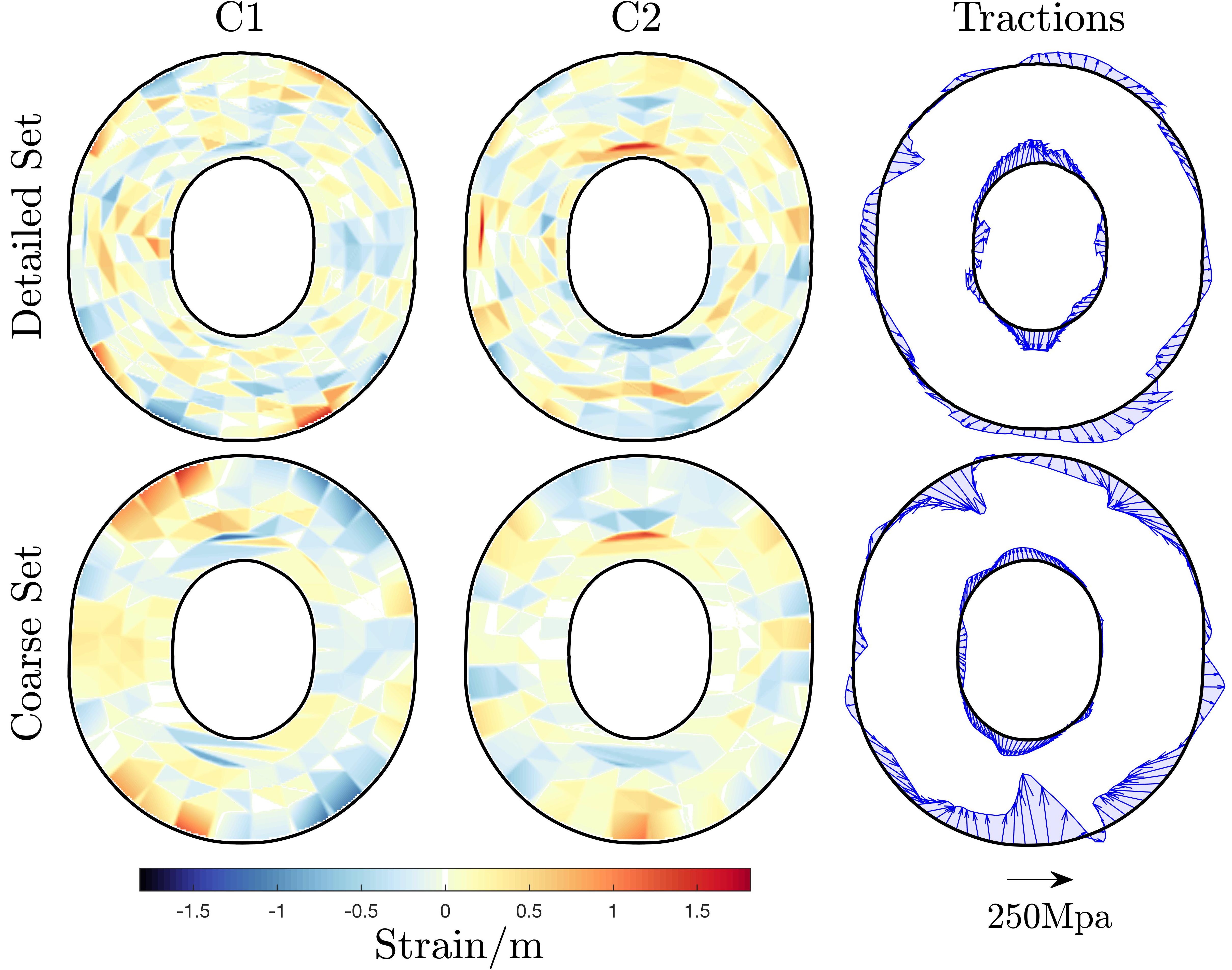}
    \caption{Violation of physical properties calculated from strain fields generated using natural neighbour interpolation with linear extrapolation. (Top and Centre) Equilibrium constraint violations. (Bottom) Equivalent tractions calculated on unloaded surfaces.}
    \label{fig:CRing_consViol}
\end{center}
\end{figure*} 

The violation of constraints calculated from \eqref{eq:equi_constraints} and \eqref{eq:traction_map} for the interpolated strain field are shown in Figure~\ref{fig:CRing_consViol}. 
The equivalent tractions have a maximum magnitude in excess of $\SI{240}{\mega\pascal}$ for detailed measurement set and $\SI{360}{\mega\pascal}$ for the coarse measurement set.
In contrast the strain fields generated using the GP method automatically satisfy both these conditions.

% section experimental (end)
% section results (end)

\section{CONCLUSION} % (fold)
\label{sec:conclusion}
A novel method for inferring strain fields from diffraction-based strain measurements is proposed.
These estimated strain fields implicitly satisfy equilibrium, ensuring their physical feasibility.
In contrast to traditional interpolation methods which treat the measurements as point-wise this GP based approach correctly relates the measurements to the average estimated strain within a gauge volume.
This improves the estimate accuracy when large and/or overlapping gauge volumes are used. 
Furthermore, multiple data sets using different size gauge volumes, measurement locations, and measurement uncertainty can be combined.
Additionally, including boundary tractions improves the accuracy of the method close to the sample boundaries and outside of the measurement locations.

This method was compared to strain fields generated using natural neighbour interpolation and linear extrapolation on both simulated and experimental data sets.
For these interpolated fields it was shown that the physical properties of equilibrium and zero tractions on the boundary were not satisfied.
In simulation, the GP approach demonstrated much greater accuracy from a limited data set created with large gauge volumes.
Experimentally, the coarse GP was able to achieve results with greater agreement to the detailed sets than the interpolation.
The results indicate that the GP method allows for larger gauge volumes to be used while still achieving estimates comparable to an interpolation with smaller gauge volumes, allowing for a reduction in beam-time.
For instance, in our experiment the GP method was able to provide more or less equivalent results on a data set that required less than 10\% of the beam time used for the detailed mapping.

The method presented in this paper is suitable for continuous planar biaxial strain fields under the assumption of plane stress. 
For discontinuous strain fields (i.e. the extreme case of the ring and plug) a mixture of experts model \cite{tresp2001mixtures}, conceptually similar to using a different GP for each sub-domain, could be used to provide good results. 
Inference of strain fields assuming plane strain requires only minor changes with Hooke's law under the assumption of plane strain substituted for Hooke's law under the assumption of plane stress. 
Future work will include the development of a software toolbox and the extension of this method to triaxial strain fields for which the Beltrami stress functions (as opposed to Airy) are appropriate.

% section conclusion (end)

\section{Acknowledgements}

This work is supported by the Australian Research Council through a Discovery Project Grant (DP170102324). Access KOWARI diffractometer was made possible an ANSTO Program Proposal PP6050.  Additional support from AINSE Limited was provided during the experimental work.

\appendix
\section{APPROXIMATION METHOD EXPRESSIONS}\label{app:approx_method}
The following provides more complete expressions for the equations given in Section~\ref{sec:approximation_method}.

As stated in Section~\ref{sec:GP_strainfield} the target strain field is related to the scalar Airy's stress functions via \eqref{eq:phitoeps}. 
Applying this linear transformation to the basis functions chosen for the Airy's stress function \eqref{eq:scalar_baseis} results in basis functions for the strain field that implicitly satisfy equilibrium;
\begin{equation*}
\begin{split}
    \phi_{\epsilon,j}(\mathbf{x}) = \mathcal{V}^{\mathbf{x}}\phi_j(\mathbf{x}) = \begin{bmatrix}\frac{\partial^2}{\partial y^2}-\nu\frac{\partial^2}{\partial x^2}\\ (1+\nu)\frac{\partial^2}{\partial x \partial y} \\ \frac{\partial^2}{\partial x^2}-\nu\frac{\partial^2}{\partial y^2} \end{bmatrix}\phi_j(\mathbf{x}),
\end{split}
\end{equation*}
where the second derivatives of the scalar basis functions are
\begin{equation*}
\begin{split}
    \frac{\partial^2}{\partial x^2}\phi_j(\mathbf{x}) &= \frac{-\lambda_{x,j}^2}{\sqrt{L_xL_y}}\sin(\lambda_{x,j}(x+L_x))\sin(\lambda_{y,j}(y+L_y)), \\
    \frac{\partial^2}{\partial y^2}\phi_j(\mathbf{x}) &= \frac{-\lambda_{y,j}^2}{\sqrt{L_xL_y}}\sin(\lambda_{x,j}(x+L_x))\sin(\lambda_{y,j}(y+L_y)), \\
    \frac{\partial^2}{\partial x\partial y}\phi_j(\mathbf{x}) &= \frac{\lambda_{y,j}\lambda_{x,j}}{\sqrt{L_xL_y}}\cos(\lambda_{x,j}(x+L_x))\cos(\lambda_{y,j}(y+L_y)). \\
\end{split}
\end{equation*}
Section~\ref{sec:GP_strainfield} also states that this strain field can be related to surface tractions via the linear mapping \eqref{eq:traction_map}, which can be used to determine basis functions for the traction constraints;
\begin{equation*}
\begin{split}
    \phi_{T,j}(\mathbf{x}) &= \mathcal{C} \phi_{\epsilon,j}(\mathbf{x})\\ 
    \phi_{T,j}(\mathbf{x}) &= \begin{bmatrix}
        n_1 & n_2 & 0 \\ 0 & n_1 & n_2
    \end{bmatrix}
    \frac{E}{1-\nu^2}
    \begin{bmatrix}
        1 & 0 & \nu \\
        0 & 1-\nu & 0 \\
        \nu & 0 & 1
    \end{bmatrix}\phi_{\epsilon,j}(\mathbf{x})
\end{split}
\end{equation*}

Section~\ref{sec:measurement_joint_distribution} states that the diffraction measurements can be related to the strain field via the linear mapping \eqref{eq:meaurement_map}. 
Basis functions for the measurements are given by applying this mapping to the basis functions for the strain field;
\begin{equation*}
    \phi_{d,j}(\boldsymbol{\eta}_i) = \mathcal{L}_{\boldsymbol\eta_i}^{\mathbf{x}} \phi_{\epsilon,j}(\mathbf{x}) = \frac{d_0 \bar{\boldsymbol\kappa}}{V}\int\limits_G\phi_{\epsilon,j}(\mathbf{x}) \, \mathrm{d}V
\end{equation*}
\begin{figure}
\begin{center}
    \includegraphics[width=0.95\linewidth]{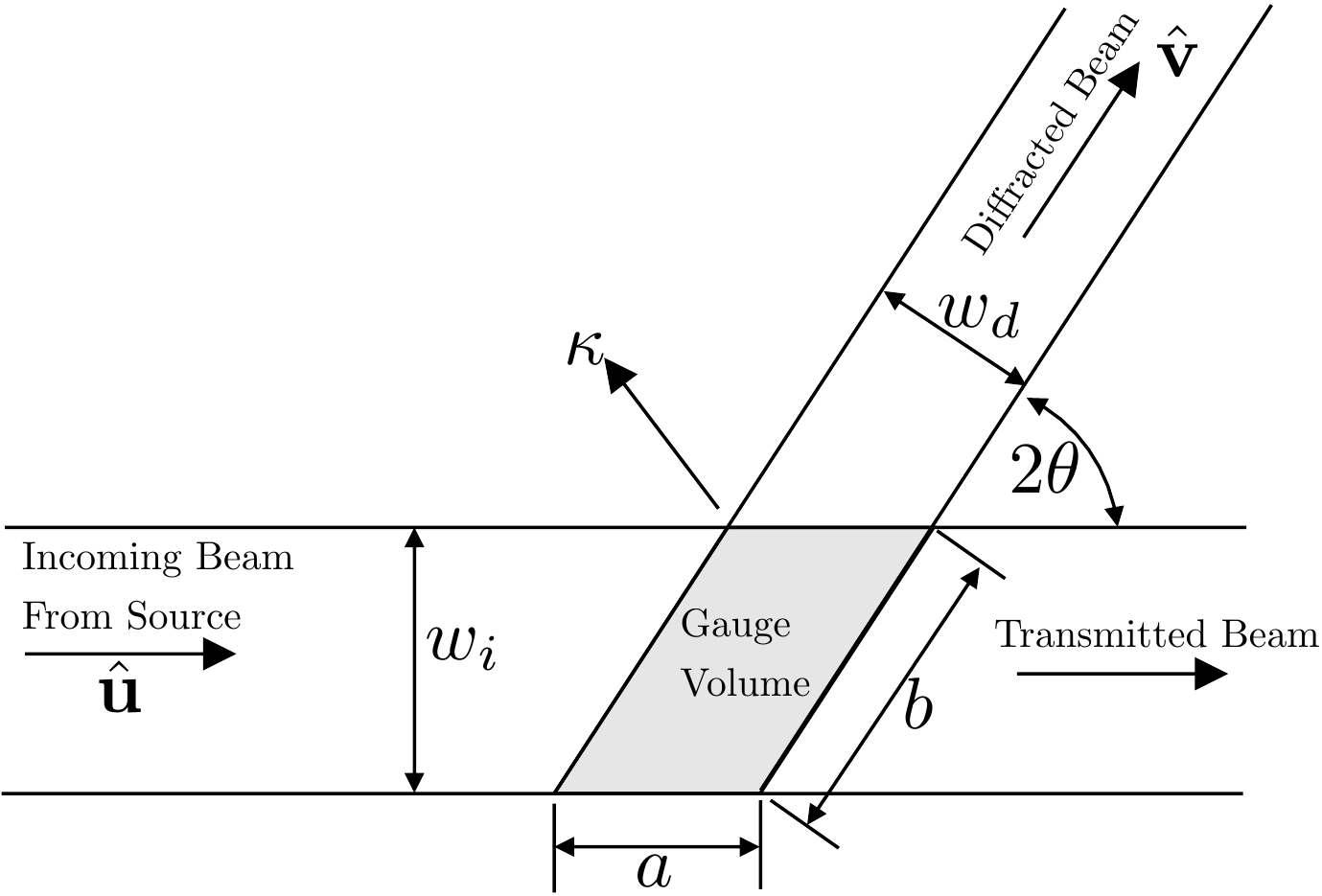}
    \caption{Gauge volume geometry for planar problem; $w_i$ is the incoming collimator slit width, $w_d$ is the diffracted beam collimator slit width, $a$ and $b$ are the side lengths of the gauge volume, $\hat{\mathbf{u}}$ is the incoming beam direction, $\hat{\mathbf{v}}$ is the diffracted beam direction, $\boldsymbol{\kappa}$ is the measurement direction, and $2\theta$ is the diffraction angle. }
    \label{fig:gauge_volume}
\end{center}
\end{figure} 

For planar problems we define the gauge volume geometry as per Figure~\ref{fig:gauge_volume}. This allows the integral over the gauge volume to be written;
\begin{equation*}
    \frac{d_0 \bar{\boldsymbol\kappa}}{V}\int\limits_G\phi_{\epsilon,j}(\mathbf{x}) \, \mathrm{d}V = \frac{d_0 \bar{\boldsymbol\kappa}}{ab \hat{\mathbf{u}}\times\hat{\mathbf{v}}}\int\limits_0^a\hspace{-2.5mm}\int\limits_0^b \phi_{\epsilon,j}(\mathbf{x}(s,q)) \,\mathrm{d}s\,\mathrm{d}q
\end{equation*}
where $\mathbf{x}(s,q) = [x_0+\hat{u}_1s+\hat{v}_1q,y_0+\hat{u}_2s+\hat{v}_2q]$ and
\begin{equation*}
\begin{split}
    &\int\limits_0^a\hspace{-2.5mm}\int\limits_0^b\frac{\partial^2}{\partial x^2}\phi_j(\mathbf{x})\,\mathrm{d}s\,\mathrm{d}q = \frac{\lambda_{x,j}^2 \left[\Gamma_1\alpha_1+\Gamma_2\alpha_2\right]}{\Gamma_3} \\
    &\int\limits_0^a\hspace{-2.5mm}\int\limits_0^b\frac{\partial^2}{\partial y^2}\phi_j(\mathbf{x})\,\mathrm{d}s\,\mathrm{d}q = \frac{\lambda_{y,j}^2 \left[\Gamma_1\alpha_1+\Gamma_2\alpha_1\right]}{\Gamma_3} \\
    &\int\limits_0^a\hspace{-2.5mm}\int\limits_0^b\frac{\partial^2}{\partial x \partial y}\phi_j(\mathbf{x})\,\mathrm{d}s\,\mathrm{d}q = \frac{-\lambda_{x,j}\lambda_{y,j} \left[\Gamma_1\alpha_2+\Gamma_2\alpha_1\right]}{\Gamma_3} \\
    &\Gamma_1 = \lambda_{x,j}\lambda_{y,j}(\hat{u}_2\hat{v}_1+\hat{u}_1\hat{v}_2)\\
    &\Gamma_2 = \lambda_{x,j}^2\hat{u}_1\hat{v}_1+\lambda_{y,j}^2\hat{u}_2\hat{v}_2 \\
    &\Gamma_3 = \sqrt{L_xL_y}(\lambda_{x,j}\hat{u}_1-\lambda_{y,j}\hat{u}_2)(\lambda_{x,j}\hat{u}_1+\lambda_{y,j}\hat{u}_2)\dots \\ 
    &\hspace{35mm} (\lambda_{x,j}\hat{v}_1-\lambda_{y,j}\hat{v}_2)(\lambda_{x,j}\hat{v}_1+\lambda_{y,j}\hat{v}_2) \\
        &\alpha_1 = \cos(\lambda_{x,j}(L_x+\hat{u}_1s+\hat{v}_1q+x_0))\dots \\
        &\hspace{35mm}\cos(\lambda_{y,j}(L_y+\hat{u}_2s+\hat{v}_2q+y_0)) \\
    &\alpha_2 = \sin(\lambda_{x,j}(L_x+\hat{u}_1s+\hat{v}_1q+x_0))\\ 
    &\hspace{35mm}\sin(\lambda_{y,j}(L_y+\hat{u}_2s+\hat{v}_2q+y_0)) \\
\end{split}
\end{equation*}

\section{EXPRESSIONS FOR PARAMETER OPTIMISATION}\label{sec:optim_expressions}
As discussed in Section~\ref{sec:parameter_optimisation} the optimal values for $\boldsymbol\theta = \{\sigma_f,l_x,l_y\}$ can be found by maximising the marginal log-likelihood $\mathcal{L}(\boldsymbol\theta)$.
The following provides the expressions to calculate the $\mathcal{L}(\boldsymbol\theta)$ and its partial derivatives when the approximation scheme presented in Section~\ref{sec:approximation_method} is used.
Letting $\bar{\mathbf{Q}} = \Phi_d\Sigma_p\Phi_d^T + \Sigma_N$ and $\bar{\mathbf{Z}} = \Phi_d^T \Sigma_N^{-1}\Phi_d + \Sigma_p^{-1}$
\begin{equation*}
\begin{split}
    &\mathcal{L}(\boldsymbol\theta) = -\frac{1}{2}\log(\bar{\mathbf{Q}}) - \frac{1}{2}\Delta\mathbf{d}^T\bar{\mathbf{Q}}^{-1} \Delta\mathbf{d}, \\
    &\log(\bar{\mathbf{Q}}) = \sum_{j=1}^{m}\log(\Sigma_{p,jj}) + \sum_{i=1}^{N}\log(\Sigma_{N,ii}) + \log(\det(\bar{\mathbf{Z}})), \\
    &\Delta\mathbf{d}^T\bar{\mathbf{Q}}^{-1} \Delta\mathbf{d} = \Delta\mathbf{d}^T\Sigma_n^{-1}\Delta\mathbf{d} - \Delta\mathbf{d}^T\Sigma_n^{-1}\Phi_d\bar{\mathbf{Z}}^{-1}\Phi_d^T\Sigma_n^{-1}\Delta\mathbf{d},
\end{split}
\end{equation*}
where $\Delta\mathbf{d} = \mathbf{d}-d_0$ and the derivatives with respect to $\theta_k$ are
\begin{equation*}
\begin{split}
    &\frac{\partial\mathcal{L}(\boldsymbol\theta) }{\partial\theta_k} = -\frac{\partial \log(\bar{\mathbf{Q}})}{\partial \theta_k} -\frac{\partial \Delta\mathbf{d}^T\bar{\mathbf{Q}}^{-1}\Delta\mathbf{d}}{\partial \theta_k} \\
    &\frac{\partial \log(\bar{\mathbf{Q}})}{\partial \theta_k} = \sum_{j=1}^{m}\left(\Sigma_{p,jj}^{-1}\frac{\partial \Sigma_{p,jj}}{\partial\theta_k}\right) - \text{trace}\left(\bar{\mathbf{Z}}^{-1}\Sigma_{p}^{-2}\frac{\partial \Sigma_{p}}{\partial\theta_k}\right)\\
    &\frac{\partial \Delta\mathbf{d}^T\bar{\mathbf{Q}}^{-1}\Delta\mathbf{d}}{\partial \theta_k} = -\Delta\mathbf{d}^T\Sigma_n^{-1}\Phi_d\bar{\mathbf{Z}}^{-1}\left(\Sigma_p^{-2}\frac{\partial\Sigma_p}{\partial\theta_k}\right)\bar{\mathbf{Z}}^{-1}\Phi_d^T\Sigma_n^{-1}\Delta\mathbf{d}
\end{split}
\end{equation*}
Algorithm~\ref{alg:optim} provides psuedo-code for a robust implementation of these expressions.

\begin{algorithm}[h!]
    \SetKwInOut{Input}{Input}
    \SetKwInOut{Output}{Output}

    % \underline{function Euclid} $(a,b)$\;
    \Input{$\Phi_*$(Estimation basis functions),$\Delta\mathbf{d}$ (measured difference), $\Phi_d$ (Measurement basis functions), $\Sigma_n$ (measurement variance), $\Sigma_p$ (basis spectral densities),$\frac{\partial\Sigma_p}{\partial\theta_k}$ (partial derivatives)} 
    $A := \begin{bmatrix}
        \Sigma_n^{-\frac{1}{2}}\Phi_d(\mathbf{x}) \\
        \Sigma_p^{-\frac{1}{2}}
    \end{bmatrix}$ \\
    $R : = \text{qr}(A)$\\
    $C := R(\text{1:$m$,1:$m$})$ \\
    $\mathbf{\alpha} := C \backslash (C^T\backslash (\Phi_d(\mathbf{x})^T\Sigma_n^{-1}\Delta\mathbf{d}))$\\
    $\log(\bar{\mathbf{Q}}) := \sum_{j=1}^{m}\log(\Sigma_{p,jj}) + \sum_{i=1}^{N}\log(\Sigma_{N,ii}) + 2\sum_{j=1}^{m}\log(|C_{jj}|) $ \\
    $\Delta\mathbf{d}^T\bar{\mathbf{Q}}^{-1}\Delta\mathbf{d} := \Delta\mathbf{d}^T\Sigma_n^{-1}\Delta\mathbf{d} - \Delta\mathbf{d}^T\Sigma_n^{-1}\Phi_d \alpha$ \\
    $\log\mathcal{L} := -\frac{1}{2}\log(\bar{\mathbf{Q}}) - \frac{1}{2}\Delta\mathbf{d}^T\bar{\mathbf{Q}}^{-1} \Delta\mathbf{d}$ \\
    \For{$k\gets0$ \KwTo $\text{length}(\theta)$}{
        $\frac{\partial \log(\bar{\mathbf{Q}})}{\partial \theta_k} := \sum_{j=1}^{m}\left(\Sigma_{p,jj}^{-1}\frac{\partial \Sigma_{p,jj}}{\partial\theta_k}\right) - \text{trace}\left(C\backslash C^T\backslash\left(\Sigma_{p}^{-2}\frac{\partial \Sigma_{p}}{\partial\theta_k}\right)\right)$ \\
        $\frac{\partial \Delta\mathbf{d}^T\bar{\mathbf{Q}}^{-1}\Delta\mathbf{d}}{\partial \theta_k} := -\alpha^T \left(\Sigma_p^{-2}\frac{\partial\Sigma_p}{\partial\theta_k}\right)\alpha$ \\
        $\frac{\partial\mathcal{L}(\boldsymbol\theta) }{\partial\theta_k} = -\frac{\partial \log(\bar{\mathbf{Q}})}{\partial \theta_k} -\frac{\partial \Delta\mathbf{d}^T\bar{\mathbf{Q}}^{-1}\Delta\mathbf{d}}{\partial \theta_k}$
    }
    \caption{log-likelihood and derivatives}\label{alg:optim}
\end{algorithm}

\FloatBarrier

% \clearpage

\section*{References}

\bibliography{References}

\end{document}